\documentclass[fleqn,12pt,twoside]{article}
\usepackage[headings]{espcrc1}

\readRCS
$Id: espcrc1.tex,v 1.2 2004/02/24 11:22:11 spepping Exp $
\usepackage{graphicx}
\usepackage[figuresright]{rotating}

\newcommand{\beq}{\begin{eqnarray}}
\newcommand{\eeq}{\end{eqnarray}}

\def\simle{\mathrel{\rlap{\raise 0.511ex \hbox{$<$}}{\lower 0.511ex \hbox{$\sim$}}}}
\def\simge{\mathrel{ \rlap{\raise 0.511ex \hbox{$>$}}{\lower 0.511ex \hbox{$\sim$}}}}

\newcommand{\ttbs}{\char'134}
\newcommand{\AmS}{{\protect\the\textfont2
  A\kern-.1667em\lower.5ex\hbox{M}\kern-.125emS}}

\title{Phases of hot and dense QCD matter}

\author{Jean-Paul Blaizot \address{Institute of Theoretical Physics, 
        CEA Saclay, \\ 
        91191 Gif-sur-Yvette cedex, France}%
        \thanks{Member of CNRS} }
       
\runtitle{Phases of hot and dense QCD matter}
\runauthor{Jean-Paul Blaizot}

\begin{document}

\maketitle

\begin{abstract}
This talk summarizes some of the theoretical issues currently most debated in the field of hot and dense QCD matter and ultra-relativistic heavy ion collisions. 
\end{abstract}

\section{Introduction}

Ultra-relativistic heavy ion collisions allow us to explore the phase diagram of matter at the highest temperature and density  that can be reached in the laboratory, and study fundamental aspects of Quantum Chromodynamics (QCD). In these extreme conditions, new phases of matter are expected to be produced.
\subsection{QCD}

By QCD matter, we mean matter where the dominant degrees of freedom are quarks and gluons, and their interactions are governed by Quantum Chromodynamics. QCD is a theory without any free parameter, except for a unique energy scale, called $\Lambda_{QCD}\approx 300$ MeV: all observables are measured in terms of (powers of) $\Lambda_{QCD}$. 
Quantities such as quark masses are fixed in nature, outside QCD, but it is sometimes convenient to allow them to vary in order to get different perspectives on the theory. Thus, for instance, chiral symmetry emerges when we let the quark masses go to zero, and many consequences of chiral symmetry, and its spontaneous breaking, remain approximately valid when the quark masses are non zero but small. In the same spirit, one can also consider analogs of QCD in which  the number of colors, or the number of flavors, go to infinity. Recently, a new playground has emerged in which some gauge theories can be solved exactly in the regime of strong coupling because they are then equivalent to a weakly coupled gravity theory. All these``variations around QCD''  provide useful indications on how QCD works. However, from the experimental point of view, we have fewer  control parameters at our disposal: essentially  the temperature and the baryon chemical potential, which can be varied by changing the  energy and the centrality of the collisions. We shall discuss the QCD phase diagram in terms of these parameters only. 

\subsection{The QCD phase diagram}

\begin{figure}
\includegraphics[scale=0.4]{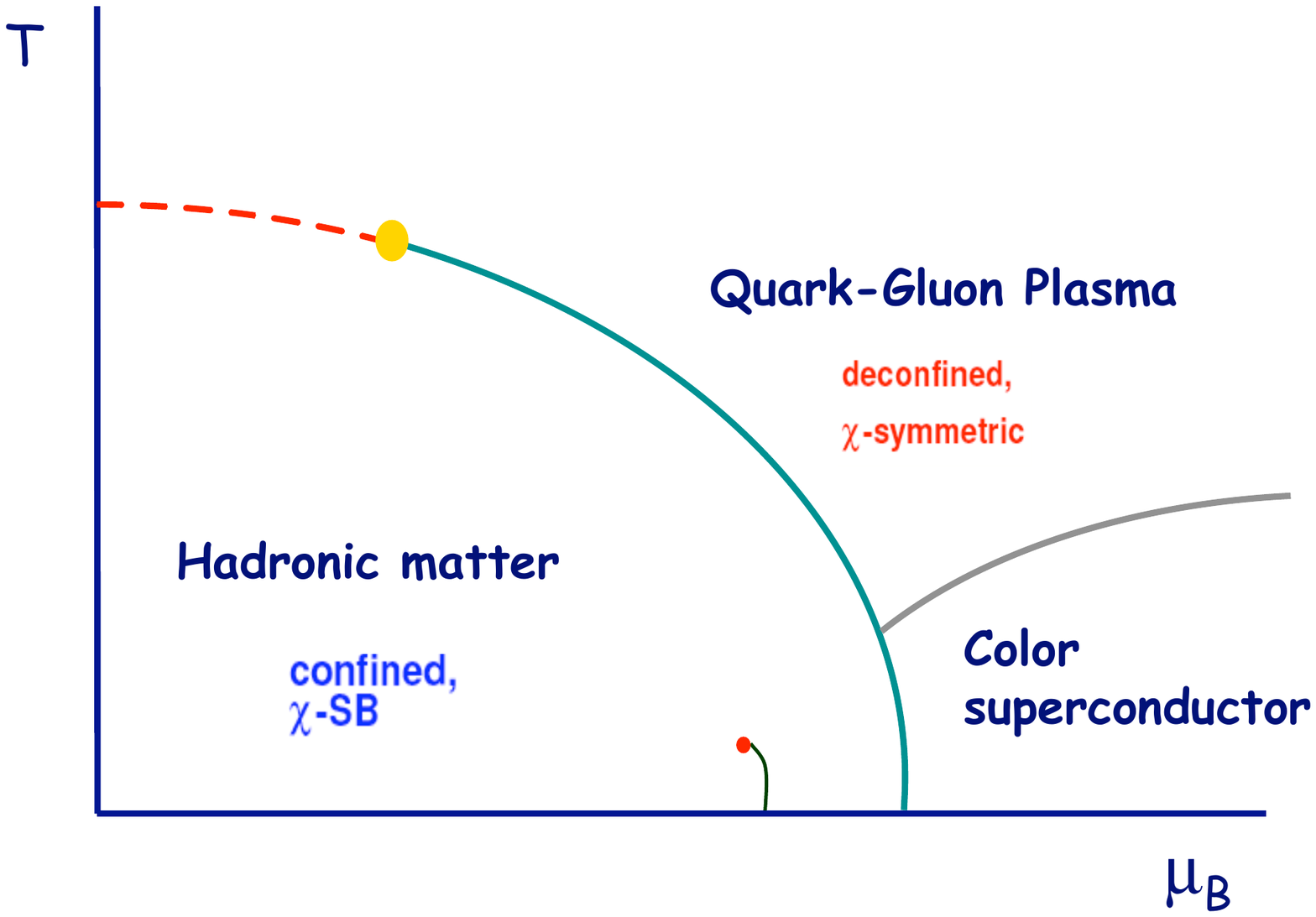}
\caption{A schematic representation of the expected phase diagram of QCD matter in the plane temperature ($T$), baryon chemical potential ($\mu_B$).}
\label{phasediagram}
\end{figure}

A ``low resolution''  phase diagram is displayed in Fig.~\ref{phasediagram}.  It  should be emphasized that this diagram represents more what we expect on the basis of various models rather than what we know.  

A basic property of QCD is the confinement of color charges: at low density and temperature quarks and gluons combine into color singlet hadrons that make up hadronic or nuclear matter.  However when the density, or the temperature,  become high enough quarks and gluons start to play a dominant role in the thermodynamics. One then expects a transition to a world dominated by quarks and gluons, and  where color is ``deconfined''. This transition is indicated by the line in the phase diagram of Fig.~\ref{phasediagram} that crosses the two axis respectively at $T_c \sim  \Lambda_{QCD}$, and  $\mu_c\sim M_N$, where $M_N$ is the nucleon mass. 

Another basic property of QCD alluded to already is chiral symmetry (an exact symmetry when quark masses vanish). Chiral symmetry  is spontaneously broken in the hadronic world, and is  expected to be restored at high temperature and density. 

\section{The ideal baryonless quark-gluon plasma}

There are at least two good reasons to focus on the case of baryon free matter: i) the baryonless quark-gluon plasma is that for which we can do the most elaborate calculations from first principles, using in particular lattice gauge theory; ii) this is likely the state of matter created in the early stages of nucleus-nucleus collisions. 

\subsection{The QCD asymptotic freedom}

QCD is ``asymptotically free'', which means that the interactions between quarks and gluons become weak when the typical energy scale ($Q$) involved is large compared to $\Lambda_{QCD}$. The strong coupling constant ``runs'', according to the (one-loop) formula
\beq
\alpha_s=\frac{g^2}{4\pi}\approx \frac{2\pi}{b_0\ln(Q/\Lambda_{QCD})}.
\eeq
Because at high temperature $Q\simeq 2\pi T$, this formula leads us to expect that 
matter becomes then ÇÊsimpleÊÈ: it turns into an ideal gas of quarks and gluons, the dominant effect of interactions being to turn (massless) quarks and gluons into weakly interacting (massive) quasiparticles. This is confirmed by weak coupling calculations (based on resummed QCD perturbation theory) that reproduce lattice results for temperatures greater than 2.5 to 3 $T_c$  \cite{Blaizot:2000fc}.  While genuine non perturbative effects may manifest themselves in specific long range correlations,  these  do not seem to affect significantly  the  bulk of  thermodynamic functions such as the pressure, the entropy or the energy density, which all go to their corresponding Stefan-Boltzmann values at high temperature. This is confirmed by new lattice calculations that can probe arbitrarily large temperatures, and which demonstrate the approach to the Stefan-Boltzmann  limit in a convincing way, in good agreement with weak coupling calculations \cite{Endrodi:2007tq}. 
The calculations of the fluctuations of conserved charges (such as baryon number, electric charge, strangeness) provide another evidence that the bulk quark behavior resembles that  a free gas  above the deconfinement transition \cite{Cheng:2008zh}.

\subsection{The cross-over between hadronic matter and the quark-gluon plasma}

Most recent lattice calculations indicate that the transition from the hadronic world to the quark gluon plasma is not a phase transition proper, but a smooth crossover \cite{Fodor:2007sy,Bazavov:2009zn}. This implies in particular that there is no unique way to define the transition temperature: it depends somewhat on how it is measured. Thus one may define the chiral transition temperature  as the location of the peak in the chiral susceptibility, and this may differ from the deconfinement temperature measured for instance by the inflexion point in the Polyakov loop expectation value. Independently of this basic ambiguity, a  discrepancy remains between the  most recent estimates of the critical temperature \cite{Fodor:2007sy,Bazavov:2009zn}. 

Between $T_c$ and  $\sim  3T_c$, there is a significant deviation between the energy density $\epsilon$, and $3P$, where $P$ is the pressure. The quantity $\epsilon-3P$, which equals the trace of the energy momentum tensor, would vanish if it were not for the fact that the QCD coupling runs, that is, it depends on the temperature. The finite value of $\epsilon-3P$ is  related to the so-called QCD scale anomaly, and  is appreciable only for $T\simle 3T_c$. 

\section{From the ``ideal gas''  to the ``perfect liquid''}

The region between $T_c$ and $ 3T_c$, is a difficult region where the physics is not well understood, but for which  much theoretical effort is needed since this is presumably the region where the quark-gluon plasma produced at RHIC spends most of its existence.  Among the  important open questions, one concerns the fate, in this region, of the quasiparticles that dominate the thermodynamics at higher temperature.  

 We shall examine some of the RHIC results, focussing  on a few which suggest in the most convincing way that the produced matter is strongly interacting, rather than exhibiting the ideal  gas behavior that would be expected if the temperature was high enough.

\subsection{Matter is opaque to the propagation of jets}

This is seen in  several ways. First by looking at the correlations, and seeing that in most central Au-Au collisions, the usual companion of a jet, expected at 180 degrees from the trigger jet, is absent \cite{Adams:2003im}.
Another view of the same physics is obtained by studying the so-called nuclear modification factor, a ratio that summarizes the deviation from what would be obtained if the nucleus-nucleus collision was an incoherent superposition of nucleon-nucleon collisions. The attenuation which persists at fairly  large transverse momentum is usually discussed in terms of the energy loss of the leading parton in the dense medium \cite{Akiba:2005bs}. This energy loss is found to be large, and difficult to account in a perturbative scheme (see e.g.  \cite{Majumder:2007iu}
 for a recent discussion). 

\subsection{Matter flows like a fluid}

If nucleus-nucleus collisions were simple superpositions of nucleon-nucleon collisions,  the produced particles
would have isotropic distributions, irrespective of the shape of the collision zone in the transverse plane. However, if the interactions among the produced particles are sufficiently strong to bring the system close to local equilibrium, then a collective motion emerges: strong pressure gradients are induced by the anisotropy of the initial interaction zone, leading to anisotropic momentum distributions\cite{Ollitrault:1992bk}. This so-called elliptic flow effect has been beautifully confirmed at RHIC, and provides strong evidence for collective behavior and (at least partial) thermalization of the produced matter.  

\subsection{The quark-gluon plasma as a perfect fluid}

The hydrodynamical calculations that are used to analyze the flow data require a short equilibration time and a relative low viscosity, i.e. a ratio of  viscosity to entropy density lower than 0.4 \cite{Luzum:2008cw}. Such a low value points to the fact that matter is strongly interacting, since the ratio of viscosity to entropy density would be much larger in a weakly interacting system. And indeed the ``measured'' value is not too different from that  obtained in some gauge theories that can be solved exactly at strong coupling: $\eta/s=1/4\pi$ \cite{Policastro:2001yc}. This latter value has in fact been conjectured to be a lower bound  \cite{Kovtun:2004de}, and no substance in nature has been found with a lower value of $\eta/s$. The small value of  $\eta/s$ obtained  for the quark-gluon plasma found at RHIC  has motivated its qualification as a ``perfect liquid''. 

\section{Is the quark-gluon plasma strongly coupled ?}

The opacity of matter, the elliptic flow, and the small value of $\eta/s$, are measurements that contribute to build a picture of the quark-gluon plasma as a strongly coupled system.

\subsection{The ideal strongly coupled quark-gluon plasma}

In fact, the RHIC data have produced a complete shift of paradigm in the field, suggesting a new ideal system that can be used as a reference system: the strongly coupled quark-gluon plasma (sQGP). This has been made possible by a  theoretical breakthrough, already alluded to, that allows one to perform  calculations in some strongly coupled gauge theories, using the so-called AdS/CFT correspondence, a mapping between a strongly coupled gauge theory 
and a weakly coupled (i.e. classical) gravity theory. This correspondence has led to the detailed calculations of many properties of  strongly coupled non  abelian 
plasmas (for a recent review see \cite{Gubser:2009fc}).
Among the successes of this approach, let us recall the exact results for the entropy density
$
{S}/{S_0}={3}/{4},
$
and  for the viscosity to entropy density ratio
$
{\eta}/{s}={1}/{4\pi}.
$

\subsection{A puzzling situation: 
weakly or strongly coupled ?  
}

The interpretation of  RHIC data in terms of a strongly coupled quark-gluon plasma leads to a somewhat puzzling situation. There is indeed no evidence that in the transition region the QCD coupling constant becomes so huge that weak coupling techniques (with appropriate resummations) are meaningless. And we know that for temperatures above $3T_c$ such calculations account well for lattice data. Besides, the description of the early stages of nucleus-nucleus collisions in terms of the color glass condensate (see below) relies heavily on weak coupling concepts. 

In fact, the situation is complicated by the coexistence within the quark-gluon plasma of degrees of freedom with different wavelengths, and whether these degrees of freedom are weakly or strongly coupled depends crucially on their wavelength. It is worth recalling here that non perturbative features may arise in a system from the 
cooperation of many degrees of freedom, or 
strong classical fields, making the system  strongly interacting while the elementary coupling strength remains small.
An illustration is provided next. 
\subsection{High density partonic systems}

To understand how the  quark-gluon plasma is produced in heavy ion collisions requires knowledge of the wave function of a nucleus at very high energy, and the detailed  mechanisms by which its partonic degrees of freedom get liberated by the collisions and subsequently interact to lead possibly to a thermalized system. 

Recall that the wave function of a relativistic system does not describe a fixed number of constituents, but rather a collection of partons, mostly gluons, whose number grows with the energy of the system: this is because each gluon acts as a color source that can radiate other gluons when the system is boosted to higher energy. When the occupation of the low (longitudinal) momentum gluon modes becomes large, a description of the wave function in terms of classical fields becomes possible. 

The gluon occupation numbers cannot grow for ever as the energy increases: when the gluons start to overlap with each other, repulsive interactions lead to ``saturation''. There is a characteristic momentum scale, called the saturation momentum $Q_s$, which grows with energy, below which all momentum states have occupation numbers of order $1/\alpha_s$. The partons with momenta larger than $Q_s$ form a dilute system. At saturation, naive perturbation theory breaks down, even though the coupling constant, whose magnitude decreases with increasing $Q_s$, may be small if $Q_s$ is large: the saturation regime is a regime of weak coupling, but large density and strong classical color fields (for a review see \cite{CGCreview}).

\section{Exploring the $\mu,T$ plane}

When moving to finite baryon chemical potential, new features appear in the phase diagram, which presents a  rich structure that is yet largely to be explored (for a recent review see \cite{BraunMunzinger:2008tz}). Among the salient features, let us mention the emergence of color supraconductivity at large density, the possible existence of a critical point, as well a a possible new phase of ``quarkyonic'' matter whose existence has been predicted on the basis of large $N_c$ arguments \cite{McLerran:2007qj}.

To guide the experimental effort in the exploration of the phase diagram, an important observation is that matter at freeze out appears to be in thermal equilibrium \cite{BraunMunzinger:2003zd}.  The simple dependence of the freeze-out parameters on the beam energy allows us to select the best experimental conditions for studying dense matter (for a recent discussion of this point, see \cite{Randrup:2006uz}). 

\section{Conclusion}

  The field has witnessed many exciting developments in recent years, both experimentally  and theoretically, and it  has a bright future since several facilities will allow us to continue our exploration of the phase diagram of hot and dense matter.  In particular the Large Hadon Collider will soon provide data that will help clarifying  many open questions and puzzles, and certainly bring us new challenges.

\end{document}